# 行为决策的阶梯理论

陈星光

（江汉大学商学院，Email: cxg@nju.edu.cn）

**摘要**：个体行为决策一个重要的特征是决策主体对多个决策属性存在一个重要性的等级分类，决策者通常是根据属性的重要性等级来确定最终的决策方案，这一点在现有的行为决策理论中并没有得到很好的考虑。本文将决策者的主观属性偏好和决策方案制定的全过程进行有机结合，对决策者在面临多属性决策任务时的决策过程进行了较为完整的描述,归纳了个体决策的方案收集和搜索寻优机制，试图揭示决策者对决策属性主观心理感知变化对于决策行为的作用机制，在此基础上，提出了行为决策的阶梯理论。通过实际案例的对比分析显示，LT 理论比两种经典的行为决策理论：前景理论（Prospect Theory）和图像理论（Image Theory）在某些决策情景下具有更好的解释和预测能力，因而具有更广泛的普适性。本文所建立的阶梯理论是对经典行为决策理论的丰富和发展，对解释现实中大量的决策现象，更好地揭示个体决策规律具有积极的理论价值和现实意义。

**关键词**：行为决策，决策理论，前景理论，图像理论，阶梯理论

## 1. 引言

自有人类产生以来，就存在着决策，决策科学是研究人类如何判断和选择的科学。在决策科学的发展历程中，大致存在着两种具有代表性的研究范式：标准化范式和描述性范式。标准化范式的基础是理性决策理论，该理论的前提假设是决策者具有完全理性，掌握完全信息，具有完备的知识，能够做出使得自己利益最大化的最优决策。标准化范式以期望值理论（ET）和主观期望效用理论（SEU）为代表，ET 和 SEU 理论都将决策视为模拟的赌博，它们的本质都源于决策者期望最大化的观念，虽然在解释数据方面，这两种观点都很有效，但是它们并没有很好地描述出真实的决策过程。后来的许多实证研究（实验室研究或者来自对现实决策的观察）对这两种基于期望最大化的决策理论提出了质疑,有些研究发现，人们在做决策时，很少考虑到显著的成本和收益均衡，Donaldson 和 Lorsch [1] 对 12 个公司展开研究发现，企业经营者并不是致力于增加股东的财富，他们的首要任务是保证公司本身的生存。

由于观察到与实际现象的不一致，有学者对 ET 和 SEU 理论进行了修正，其中最为著名的要数 Kahneman 和 Tversky [2,3] 提出的前景理论（Prospect Theory，PT）以及后来的累积前景理论（Cumulative Prospect Theory， CPT），前景理论保留了一般期望值最大化的假设，加入了决策者的心理假设，使得理论对于实际决策行为更具有预测性，从而成为决策科学中描述型范式的杰出代表。随着行为决策理论的进一步发展，现实中的决策行为出现了现有理论无法解释的情况，主要的问题来自三点：（1）多数情况下，大多数人对于大多数问题使用某种简单、易用、快速、无须分析的过程（实证研究显示，在许多决策情形中，概率对许多人并没有多大意义，以往的基于赌博模拟的理论，比如 ET、SEU、PT、CPT 中，对于概率的运用过于频繁，实际决策行为概率的作用往往很模糊，并不明显）；（2）决策的选择方案集合是如何确定下来的？（3）决策者经常选择那些很明显不符合他们最大利益的选项，现实中，决策制定受到信念、道德、伦理和社会公约的强烈影响。

为了克服这些缺陷，Beach 和 Mitchell [4,5] 构建了行为决策的图像理论

（Image theory，IT），通过引入一个反映决策者关于良好而合适状态的设想——图像（image），来描述实际中决策的制定和选择过程，该理论将决策视为一组相互联系的原则、目标和计划，决策者心目中有一组潜在理想的"图像"，决策通过兼容性测试和收益性测试机制进行展开，通过兼容性测试得到决策的方案集，通过收益性测试得到最终的决策方案，一系列实证研究支持了该理论的主要观点，从而较好地解决了上面三个冲突。图像理论的问题在于，对收益性测试阶段的描述过于简化，该理论主要从数量的角度进行最终方案的筛选，而现实中的决策问题可选择的方案往往具有多个属性，图像理论对于这种多属性决策的情形考虑得不够充分，导致对有些决策问题的实际解释力不够，在本文后面的案例分析中对此现象有进一步的阐述。

本文在前人研究的基础上，试图从行为决策的角度，针对现有理论的不足进行改进，事实上，通过对日常决策行为的观察，可以发现在现实中，我们是通过比较不同方案之间优势而获得最终的决策方案的，并且，实际起作用的往往是那些对决策个体而言最重要的决策属性（可能不止一个），如果某个方案在这些最重要的决策属性上占有明显优势，决策者往往会最终选择该方案。其次，属性的重要程度往往是随着决策进程的发展而发生变化的，基于这些经验事实，我们提出了个体行为决策的阶梯理论（Ladder Theory，LT），该理论能够涵盖行为决策理论中的 PT 和 IT 两大经典理论，克服了基于 PT 理论过于依赖概率而与某些实际决策行为偏离的问题，同时对于多属性决策问题可以提供较好的解释。

本文第一部分是引言，介绍研究背景并进行文献回顾，第二部分介绍阶梯理论的基本模型，包括前提假设、符号定义和模型框架，第三部分通过四个实际案例说明了阶梯理论的合理性和有效性，第四部分在前面分析的基础上，对阶梯理论展开进一步讨论，第五部分对全文进行了总结。

2. 阶梯理论的基本模型
2.1 假设、符号和定义
1）前提假设
阶梯理论针对的是个体决策行为，基于如下三个基本假设：
（1）决策行为的依赖性假设：决策者的行为，取决于该决策者所面临的决策情景（包括决策任务和决策环境），不同的决策情景决定了决策者对于相同决策属性的重要度偏好不同，即决策属性的重要度偏好是依赖于决策情景的；
（2）决策任务的多属性假设：对于某个特定决策者的某项决策任务，具有一系列决策属性，各决策属性对于决策者而言，具有特定的重要性；
（3）决策方案的可辨识性假设：对于某个特定决策者的决策行为，不存在两个完全相等的决策方案，即决策者待选择的任意两个方案不会完全相同。
2）符号说明
（1）$D_i^m$：某个决策者 $i$ 在决策任务 $m$ 上的一个决策行为；

（2）$\mathbf{S}(D_i^m) := \{s_{i,1}^m, s_{i,2}^m, \cdots, s_{i,|\mathbf{S}(D_i^m)|}^m\}$：决策行为 $D_i^m$ 中所有可能的决策方案集合，集合中元素的个数为 $|\mathbf{S}(D_i^m)|$，每一个元素 $s_{i,j}^m$ 表示决策者 $i$ 在决策任务 $m$ 上的第 $j$ 个方案；

（3）$\tilde{\mathbf{S}}(D_i^m) := \{\tilde{s}_{i,j}^m : \tilde{s}_{i,j}^m \in \mathbf{S}(D_i^m), j \in \{1, \cdots, |\tilde{\mathbf{S}}(D_i^m)|\}\} \subseteq \mathbf{S}(D_i^m)$：决策行为 $D_i^m$ 的实际

可行选择方案集合，是所有可能决策方案集合 $\mathbf{S}_i^m$ 的子集；

（4）$\bar{\mathbf{S}}(D_i^m) := \{\bar{s}_{i,j}^m : \bar{s}_{i,j}^m \in \tilde{\mathbf{S}}(D_i^m), j \in \{1, \cdots, |\bar{\mathbf{S}}(D_i^m)|\}\} \subseteq \tilde{\mathbf{S}}(D_i^m)$：决策行为 $D_i^m$ 的最终决策方案集合，是实际可行选择方案集合 $\tilde{\mathbf{S}}(D_i^m)$ 的子集；

（5）$\mathbf{A}(D_i^m) = \{1, 2, \cdots, N(D_i^m)\}$：决策行为 $D_i^m$ 上的属性集合，每一个元素代表决策者在决策中所考虑的一个属性，$N(D_i^m)$ 是行为 $D_i^m$ 上的所有属性的数量；

（6）$\mathbf{A}_\alpha(D_i^m) = \{\alpha_1, \alpha_2, \cdots, \alpha_{N_\alpha(D_i^m)}\} \subseteq \mathbf{A}(D_i^m)$：决策行为 $D_i^m$ 上的基本属性集合，表示在实际可行方案形成阶段对决策者起作用的属性，集合中元素的个数 $|\mathbf{A}_\alpha(D_i^m)| = N_\alpha(D_i^m)$，由决策行为 $D_i^m$ 确定，$N_\alpha(D_i^m)$ 表示决策行为 $D_i^m$ 基本属性的数量；

（7）$\mathbf{A}_\beta(D_i^m) = \{\beta_1, \beta_2, \cdots, \beta_{N_\beta(D_i^m)}\} \subseteq \mathbf{A}(D_i^m)$：决策行为 $D_i^m$ 上的占优属性集合，表示在最终方案形成的阶段对决策者起作用的属性，$N_\beta(D_i^m)$ 表示决策行为 $D_i^m$ 占优属性的数量；（注意，$\mathbf{A}_\alpha(D_i^m)$、$\mathbf{A}_\beta(D_i^m)$ 和 $\mathbf{A}(D_i^m)$ 可能完全相同）

存在关系：$\mathbf{A}(D_i^m) = \mathbf{A}_\alpha(D_i^m) \bigcup \mathbf{A}_\beta(D_i^m)$，显然，$N_\alpha(D_i^m) \leq N(D_i^m)$，$N_\beta(D_i^m) \leq N(D_i^m)$，即决策中所有的属性，分为基本属性和占优属性两大类，这两类属性可能存在交叉，即，可能存在某个属性既是基本属性也是占优属性的情况；

（8）$\mathbf{x}(\mathbf{A}_\alpha(D_i^m)) := (a_{i,1}^m, \cdots, a_{i,N_\alpha(D_i^m)}^m)$：决策行为 $D_i^m$ 的基本属性价值向量，每一个元素 $a_{i,j}^m$ 表示为决策行为 $D_i^m$ 上的第 $j$ 个属性的基本值（门槛值），表示决策者对该属性价值的心理预期临界值，即某个方案对应的属性价值至少不能低于该值，否则，该方案不会被最终选择，$j \in \{1, \cdots, N_\alpha(D_i^m)\}$；

（9）$\bigcup_{j=1}^{L(D_i^m)} \mathbf{B}_j(D_i^m) = \mathbf{A}_\beta(D_i^m)$：决策行为 $D_i^m$ 上的占优属性分割，由决策者在进行决策时候根据属性的重要性预先给定，其中，$L(D_i^m)$ 表示决策行为 $D_i^m$ 的占优属性等级数量，子集 $\mathbf{B}_j(D_i^m)$ 称为占优属性集合 $\mathbf{A}_\beta(D_i^m)$ 上的第 $j$ 级属性子集，任意两个子集 $\mathbf{B}_j(D_i^m) \bigcap \mathbf{B}_k(D_i^m) = \varnothing$，$j, k \in \{1, \cdots, L(D_i^m)\}$；随着下标 $j$ 的增加，该级

别决策属性的重要性依次升高，每一个子集 $\mathbf{B}_j(D_i^m) := \{j_1, \cdots, j_{n(j)} : j_k \in \{1, \cdots, N_\beta(D_i^m)\}\}$ 表示属性代号为 $j_1, \cdots, j_{n(j)}$ 的 $n(j)$ 个决策属性，对于决策者而言，具有相同的重要性，下标满足如下关系：

$$\sum_{j=1}^{L(D_i^m)} n(j) = N_\beta(D_i^m) ;$$

（10） $\mathbf{x}(s_{i,j}^m) := (x_{i,j,1}^m, \cdots, x_{i,j,N(D_i^m)}^m)$：决策方案 $s_{i,j}^m$ 的价值向量，其中的分量 $x_{i,j,k}^m$ 表示对决策者而言，决策方案 $s_{i,j}^m$ 在第 $k$ 个属性上的价值；

（11） $\mathbf{x}_\alpha(s_{i,j}^m) := (x_{i,j,\alpha_1}^m, \cdots, x_{i,j,\alpha_{N_\alpha(D_i^m)}}^m)$：决策方案 $s_{i,j}^m$ 对应于基本属性的价值向量；

（12） $\mathbf{x}_\beta(s_{i,j}^m) := (x_{i,j,\beta_1}^m, \cdots, x_{i,j,\beta_{N_\beta(D_i^m)}}^m)$：决策方案 $s_{i,j}^m$ 对应于占优属性的价值向量；

（13） $\mathbf{X}(\mathbf{S}(D_i^m)) := (\mathbf{x}(s_{i,1}^m), \mathbf{x}(s_{i,2}^m), \cdots, \mathbf{x}(s_{i,|\mathbf{S}(D_i^m)|}^m))^T$：决策行为 $D_i^m$ 所有可能决策方案 $\mathbf{S}(D_i^m)$ 的价值矩阵；

（14） $\mathbf{X}(\tilde{\mathbf{S}}(D_i^m)) := (\mathbf{x}(\tilde{s}_{i,1}^m), \mathbf{x}(\tilde{s}_{i,2}^m), \cdots, \mathbf{x}(\tilde{s}_{i,|\tilde{\mathbf{S}}(D_i^m)|}^m))^T$：决策行为 $D_i^m$ 实际可行选择方案 $\tilde{\mathbf{S}}(D_i^m)$ 的价值矩阵；

（15） $\mathbf{X}_\alpha(\mathbf{S}(D_i^m)) := (\mathbf{x}_\alpha(s_{i,1}^m), \mathbf{x}_\alpha(s_{i,2}^m), \cdots, \mathbf{x}_\alpha(s_{i,|\mathbf{S}(D_i^m)|}^m))^T$：决策行为 $D_i^m$ 所有可能决策方案 $\mathbf{S}(D_i^m)$ 对应于基本属性的价值矩阵；

（16） $\mathbf{X}_\beta(\tilde{\mathbf{S}}(D_i^m)) := (\mathbf{x}_\beta(\tilde{s}_{i,1}^m), \mathbf{x}_\beta(\tilde{s}_{i,2}^m), \cdots, \mathbf{x}_\beta(\tilde{s}_{i,|\tilde{\mathbf{S}}(D_i^m)|}^m))^T$：决策行为 $D_i^m$ 实际可行选择方案 $\tilde{\mathbf{S}}(D_i^m)$ 对应于占优属性的价值矩阵；

（17） $\mathbf{x}_k(\tilde{\mathbf{S}}(D_i^m)) := \{x_{i,l,k}^m : l = 1, \cdots, |\tilde{\mathbf{S}}(D_i^m)|\}$：从决策行为 $D_i^m$ 实际可行选择方案集合 $\tilde{\mathbf{S}}(D_i^m)$ 属性价值向量的每个分量选择出第 $k$ 个分量组成的集合；

（18） $f_i^m$：$\mathbf{S}(D_i^m) \to \mathbf{X}(\mathbf{S}(D_i^m))$，决策行为 $D_i^m$ 上的方案—属性映射，该映射得到决策行为 $D_i^m$ 所有可能决策方案 $\mathbf{S}(D_i^m)$ 的属性价值矩阵 $\mathbf{X}(\mathbf{S}(D_i^m))$；

（19） $\hat{\mathbf{S}}(\mathbf{B}_k(D_i^m)|\mathbf{S}(D_i^m)) := \{\hat{s}_1(\mathbf{B}_k(D_i^m)|\mathbf{S}(D_i^m)), \cdots, \hat{s}_{Q(k)}(\mathbf{B}_k(D_i^m)|\mathbf{S}(D_i^m))\}$：决策方案集合 $\mathbf{S}(D_i^m)$ 在属性子集 $\mathbf{B}_k(D_i^m)$ 上的占优方案集合，集合元素的数量为 $Q(k)$。

3）定义

定义 1：属性集 $\tilde{\mathbf{A}}(D_i^m)$ 上的占优方案。对于某个决策行为 $D_i^m$ 上的方案序列 $\mathbf{S}(D_i^m)$，称方案 $\hat{s}(\tilde{\mathbf{A}}(D_i^m)|\mathbf{S}(D_i^m))$ 为方案集合 $\mathbf{S}(D_i^m)$ 在属性集 $\tilde{\mathbf{A}}(D_i^m)$ 上的占优方案，如果对于任意属性集 $\tilde{\mathbf{A}}(D_i^m)$ 对应的 $|\tilde{\mathbf{A}}(D_i^m)|$ 个属性，方案 $s_{i,j}^m$ 至少有一个属性价值严格优于其它 $|\mathbf{S}(D_i^m)|-1$ 个方案的该属性价值，并且其它的 $|\tilde{\mathbf{A}}(D_i^m)|-1$ 个属性价值不劣于其它 $|\mathbf{S}(D_i^m)|-1$ 个方案对应的 $|\tilde{\mathbf{A}}(D_i^m)|-1$ 个属性价值，则称方案 $s_{i,j}^m$ 为属性集 $\tilde{\mathbf{A}}(D_i^m)$ 上的占优方案。

定义 2：决策方案 $s_{i,1}^m$ 和 $s_{i,2}^m$ 完全相等。两个决策方案 $s_{i,1}^m$ 和 $s_{i,2}^m$ 称为完全相等，如果 $\mathbf{x}_\alpha(s_{i,1}^m) = \mathbf{x}_\alpha(s_{i,2}^m)$，并且 $\mathbf{x}_\beta(s_{i,1}^m) = \mathbf{x}_\beta(s_{i,2}^m)$，即决策方案 $s_{i,1}^m$ 和 $s_{i,2}^m$ 的价值向量 $\mathbf{x}(s_{i,1}^m)$ 和 $\mathbf{x}(s_{i,2}^m)$ 在基本属性集和占优属性集上的对应分量完全相等。

2.2 LT 理论的决策模型
　　LT 理论将一般的行为决策过程分为两个大的阶段，第一阶段，实际可行方案的收集阶段，第二阶段，最终待执行方案生成阶段。这里分别阐述如下：
1）LT 的第一阶段
　　对于决策行为 $D_i^m$，决策者根据决策任务 $m$ 的特征，首先自发产生一个决策任务的基本属性集 $\mathbf{A}_\alpha(D_i^m)$，同一个决策者面对不同决策任务，其基本属性集 $\mathbf{A}_\alpha(D_i^m)$ 可能不同，然后，决策者将所有可能决策方案的基本属性价值矩阵 $\mathbf{X}_\alpha(\mathbf{S}(D_i^m))$ 中的每一行（即每个决策方案的基本价值向量）与决策任务的基本属性价值向量 $\mathbf{x}(\mathbf{A}_\alpha(D_i^m))$ 相比较，如果存在某个方案的某个属性价值低于基本属性价值向量的对应属性的价值，则该方案被淘汰，否则，该方案进入选择集，通过该步骤，得到决策可行方案选择集。该步骤往往是一个内隐的过程，即在决策过程中，该过程不明显出现在决策者的决策心理活动中，而是通过潜意识进行。这里，可行选择方案集 $\tilde{\mathbf{S}}(D_i^m)$ 中不存在完全相同的两个方案，即没有方案冗余。将此阶段称为内隐筛选阶段，得到决策可行选择方案集的过程称为初步筛选过程，该过程可以形式化描述如下：
**初步筛选过程（Primary sifting process, PSP）：**
　　（1）产生决策行为 $D_i^m$ 所有可能决策方案的基本属性价值矩阵，即通过 $f_i^m$ 得到 $\mathbf{X}_\alpha(\mathbf{S}(D_i^m))$；

（2）方案筛选：比较 $\mathbf{X}_\alpha(\mathbf{S}(D_i^m))$ 中每一个方案的基本属性价值向量 $\mathbf{x}_\alpha(s_{i,j}^m)$ 与 $\mathbf{x}(\mathbf{A}_\alpha(D_i^m))$ 中对应元素的相对优劣，如果存在 $\mathbf{x}_\alpha(s_{i,j}^m)$ 的某个分量 $x_{i,j,k}^m$，使得 $x_{i,j,k}^m \prec a_{i,k}^m$，即方案 $s_{i,j}^m$ 上的第 $k$ 个基本属性的价值劣于基本属性集合 $\mathbf{A}_\alpha(D_i^m)$ 中第 $k$ 个元素的值，则决策方案 $s_{i,j}^m$ 从所有可能的决策方案集合 $\mathbf{S}(D_i^m)$ 中被剔除，这里 $j=1,\cdots,|\mathbf{S}(D_i^m)|$，$k=1,\cdots,N_\alpha(D_i^m)$；

（3）重复第（2）步，直到决策方案集合 $\mathbf{S}(D_i^m)$ 中没有方案可以被剔除为止，则得到最终的决策行为 $D_i^m$ 的实际可行选择方案集合 $\tilde{\mathbf{S}}(D_i^m) = \mathbf{S}(D_i^m)$。

2）LT 的第二阶段

其基本思想是，首先比较等级最高的属性子集 $\mathbf{B}_{N_\beta(D_i^m)}(D_i^m)$ 上所有方案的占优属性价值，从中找出属性子集 $\mathbf{B}_{N_\beta(D_i^m)}(D_i^m)$ 上的占优方案，如果存在两个或两个以上的方案，再比较次级属性价值子集 $\mathbf{B}_{N_\beta(D_i^m)-1}(D_i^m)$ 上所有方案的占优属性价值，从中找出属性子集 $\mathbf{B}_{N_\beta(D_i^m)-1}(D_i^m)$ 上的占优方案，如仅有一个方案，则该方案为最终方案。否则继续进行上述步骤，一直到找到一个方案，该方案满足在所有属性价值子集 $\mathbf{B}_j(D_i^m)$ 上均为占优方案（$j \in \{1,\cdots,L(D_i^m)\}$），则该方案即为最终的决策方案。如果在某步中不存在占优方案，则决策者重新调整决策行为 $D_i^m$ 上的占优属性分割，得到一个新的占优方案集合 $\bigcup_{j=1}^{L(D_i^m)} \mathbf{B}'_j(D_i^m)$，再继续本阶段。

特殊情形是，可行选择方案集合 $\tilde{\mathbf{S}}(D_i^m)$ 中方案的个数为一个，而该方案不能在最高等级的属性价值子集满足决策者的预期标准，则该决策者会选择放弃执行该方案。反之，如果该方案在最高等级的属性价值子集上满足决策者的预期标准，则决策者往往会选择执行该方案。具体执行与否，与决策者的个体特征密切相关。该过程可以形式化描述如下：

**阶梯搜索过程：( Ladder search process, LSP)**

（1）设置初始占优属性等级变量 $r = L(D_i^m)$，表示属性的最高等级，得到可行方案集 $\tilde{\mathbf{S}}(D_i^m)$ 在属性价值子集 $\mathbf{B}_r(D_i^m)$ 上的占优方案集合 $\hat{\mathbf{S}}(\mathbf{B}_r(D_i^m)|\tilde{\mathbf{S}}(D_i^m))$；

（2）如果占优方案集合 $\hat{\mathbf{S}}(\mathbf{B}_r(D_i^m)|\tilde{\mathbf{S}}(D_i^m))$ 中的元素数量 $Q(k) > 1$，则将属性等级

变量减少1，即 $r = r-1$，得到新的占优方案集合 $\hat{\mathbf{S}}(\mathbf{B}_r(D_i^m)|\tilde{\mathbf{S}}(D_i^m))$，重复第（2）步，直到 $Q(k)=1$，则占优方案集合 $\hat{\mathbf{S}}(\mathbf{B}_r(D_i^m)|\tilde{\mathbf{S}}(D_i^m))$ 中的唯一方案为最终决策方案，最终决策方案集合 $\bar{\mathbf{S}}(D_i^m) = \{\hat{s}_{i,1}^m : \hat{s}_{i,1}^m \in \hat{\mathbf{S}}(\mathbf{B}_r(D_i^m)|\tilde{\mathbf{S}}(D_i^m))\}$。

之所以称为阶梯理论，是一个形象的比喻，该选择机制类似于一个梯子的形状，每一次判断过程，降低一个占优属性分割的等级进行比较，直到找到一个最优的等级，比较停止，称此阶段为外显寻优阶段。阶梯搜索过程的示意图如下图1所示。

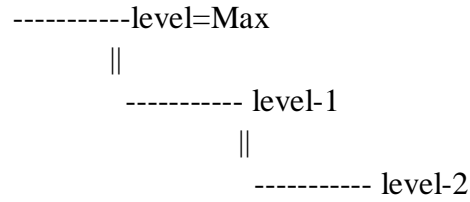

图 1　阶梯搜索示意图

上图中，梯子之间的距离，表示相邻属性等级之间的相对重要性的差距，梯子的水平长度，表示在该等级属性集合上，方案价值的最大值。

对于上述决策方案的形成机制，我们有如下命题保证最后可得到唯一的待执行方案。

命题1：阶梯搜索过程在有限步内收敛到唯一的最终方案。

证明：由阶梯搜索过程可知，阶梯搜索过程的步骤最多为 $L(D_i^m)$ 步，假设前面进行了 $L(D_i^m)-1$ 步，每一步得到的占优方案集合元素个数 $Q(k)>1$，最后一步中，属性类别等级变量 $r=1$（最低等级），则最后一步的占优方案集合 $\hat{\mathbf{S}}(\mathbf{B}_r(D_i^m)|\tilde{\mathbf{S}}(D_i^m))$ 中方案数量 $|\hat{\mathbf{S}}(\mathbf{B}_r(D_i^m)|\tilde{\mathbf{S}}(D_i^m))|$ 必然等于 1，否则，假设最后一步的占优方案集合 $\hat{\mathbf{S}}(\mathbf{B}_r(D_i^m)|\tilde{\mathbf{S}}(D_i^m))$ 中方案数量 $|\hat{\mathbf{S}}(\mathbf{B}_r(D_i^m)|\tilde{\mathbf{S}}(D_i^m))|>1$，则存在两个或两个以上最低等级的占优方案，由占优方案的定义和阶梯搜索过程 LSP 易知，在可行方案集合 $\tilde{\mathbf{S}}(D_i^m)$ 中，至少存在两个或两个以上的方案，在所有占优属性上的价值完全相等，阶梯搜索过程中，方案的所有属性都为占优属性，即在可行方案集合 $\tilde{\mathbf{S}}(D_i^m)$ 中，至少存在两个或两个以上的方案完全相等，这与阶梯理论的假设（3）矛盾，故至多经过 $L(D_i^m)$ 步阶梯搜索，得到包含一个方案的占优方案集合，该方案即为满足条件的唯一最终方案，证毕。

下面的图2描述了阶梯理论框架下一个完整的决策行为过程。

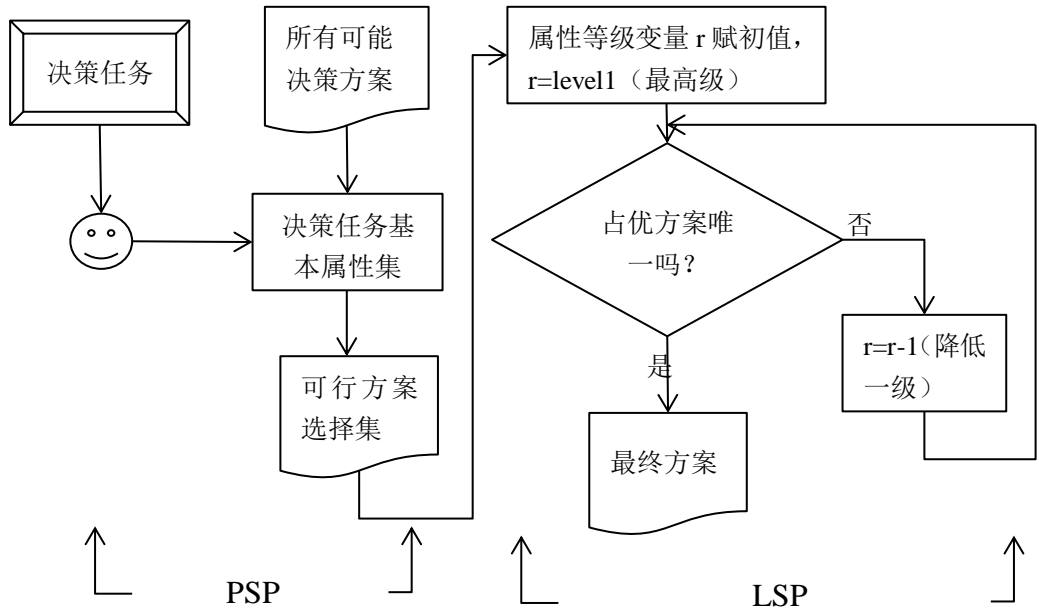

图 2 阶梯理论决策机制示意图

3. 案例分析

3.1 （案例 1）出行行为决策（出差赶车）

出行者 $i$ 需要携带行李，从 A 到 B 地，可能的出行决策方案：m1.地铁；m2. 公交车；m3.小汽车（本案例能够用 PT、IT、LT 理论解释）。

对于普通的交通出行，出行者 $i$ 的出行任务属性集为{1：时间；2：金钱；3：舒适度；4：安全性；5：误点风险}，属性 1、2、5 指标越小越好；属性 3、4 越高越好；

假设：出行者 $i$ 从 A 到 B 地，公交车的行程时间为[50,60]，小汽车的行程时间为[30,70]分钟，由于驾驶者不熟悉路况，故小汽车可能的最长行程时间比较大；

对于定性指标如安全性、舒适度、误点风险等分为 5 个等级，对应关系为：极低（very low）—1；低（low）—2；中等（moderate）—3；高（high）—4；极高（very high）—5；

实际结果：出行者由于不熟悉路况，放弃了小汽车出行方案，又由于公交车的误点风险比较大，放弃了公交车出行方案，出行者最终选择了风险最小的地铁出行方案。

（1）应用 LT 理论

出行者 $i$ 的决策任务 1（出差赶车）属性集 $\mathbf{A}(D_i^1)$ 为：

$\mathbf{A}(D_i^1) = \{1,2,3,4,5\}$

出行者 $i$ 的决策任务 1（出差赶车）的基本属性集合 $\mathbf{A}_\alpha(D_i^1)$ 为：

$\mathbf{A}_\alpha(D_i^1) = \{1,2,3,4,5\}$

出行者 $i$ 的决策任务 1（出差赶车）的占优属性集合 $\mathbf{A}_\beta(D_i^1)$ 为：

$$\mathbf{A}_\beta(D_i^1) = \{1,2,3,4,5\}$$

基本属性价值向量 $\mathbf{x}(\mathbf{A}_\alpha(D_i^1))$ 为：

$$\mathbf{x}(\mathbf{A}_\alpha(D_i^1)) = (\leq 50, \leq 100, \geq \text{"moderate"-3}, \geq \text{"high"-4}, \leq \text{"low"-2})$$

出行者 $i$ 的决策任务 1（出差赶车）的占优属性分割为：

$$\mathbf{B}_1(D_i^1) \cup \mathbf{B}_2(D_i^1) = \{1,2,3\} \cup \{4,5\}$$

出行者 $i$ 的决策任务 1 的所有可能决策方案集合：

$$\mathbf{S}(D_i^1) = \{s_{i,1}^1 = m1, s_{i,2}^1 = m2, s_{i,3}^1 = m3\}$$

依据 2.1 小节的符号和定义，通过决策行为 $D_i^1$ 上的方案—属性映射 $f_i^1$ 得到出行者 $i$ 的决策任务 1 的基本属性价值矩阵：

$$\mathbf{X}_\alpha(\mathbf{S}(D_i^1)) = (\mathbf{x}(s_{i,1}^1), \mathbf{x}(s_{i,2}^1), \mathbf{x}(s_{i,3}^1))^T = \begin{pmatrix} 40 & 50 & 3 & 5 & 1 \\ 50\sim 60 & 50 & 3 & 4 & 3 \\ 30\sim 70 & 90 & 4 & 4 & 2 \end{pmatrix}$$

出行者 $i$ 的决策任务 1 的占优属性价值矩阵 $\mathbf{X}_\beta(\mathbf{S}(D_i^1))$ 与基本属性价值矩阵 $\mathbf{X}_\alpha(\mathbf{S}(D_i^1))$ 完全相同；

运用 2.2 小节的 PSP 过程，比较 $\mathbf{X}_\alpha(\mathbf{S}(D_i^1))$ 的每一行和基本属性价值向量 $\mathbf{x}(\mathbf{A}_\alpha(D_i^1))$ 中每一个元素的相对优劣，由于出行方案 $s_{i,2}^1$ 的第 5 个属性的价值为 3，劣于基本属性价值向量 $\mathbf{x}(\mathbf{A}_\alpha(D_i^1))$ 第 5 个属性的基本值 2（出行误点风险值越小越好），因此出行方案 $s_{i,2}^1$ 被剔除，易验证其它两个方案被保留，则出行者 $i$ 的决策任务 1 的可行决策方案集合：$\tilde{\mathbf{S}}(D_i^1) = \{s_{i,1}^1 = m1, s_{i,3}^1 = m3\}$；

再应用 2.2 小节的 LSP 过程，占优属性分割：$\mathbf{A}_\beta(D_i^1) = \mathbf{B}_1(D_i^1) \cup \mathbf{B}_2(D_i^1) = \{1,2,3\} \cup \{4,5\}$，决策行为 $D_i^1$ 的占优属性等级数 $L(D_i^m) = 2$，在最高等级 $r = 2$ 上的占优方案集合 $\hat{\mathbf{S}}(\mathbf{B}_2(D_i^1) | \tilde{\mathbf{S}}(D_i^1)) = s_{i,1}^1$（方案 $s_{i,1}^1$ 和方案 $s_{i,2}^1$ 在第 4 个属性和第 5 个属性上的价值分别为 $\{5,1\}$ 和 $\{4,2\}$，明显有 $\{5,1\} \succ \{4,2\}$）。故在案例 1 中，出行者的最终出行方案为 $s_{i,1}^1$ 即选择地铁出行的方案 m1，与实际情况一致。

（2）应用 PT 理论

根据前景理论的基本假设，决策者面对获得的时候是风险厌恶的，决策者面对损失的时候是风险偏好的，由于出行行为可以看作是一种获得，因此，根据 PT 理论，决策者在选择出行方式的时候，将最小化自己的行为风险，因此将选择出行方式 1，即乘地铁出行，与 LT 理论预测的行为一致；

（3）应用 IT 理论

由于 IT 理论的收益性测试仅仅从量的最优方面考虑决策行为，因此当方案 1 和方案 2 通过 IT 理论的兼容性测试后，无法进一步从方案 1 和方案 2 中进行选择得到最终的决策方案，因此，对于本问题，IT 理论将失效。

3.2 （案例 2）出行行为决策（外出旅游）：出行者 $i$ 需从 C 到 D 地进行自助旅游，可能的决策方案： m1. 地铁；m2.公交车；m3.小汽车（本案例不能够用 PT、IT 理论解释，能够用 LT 理论解释）。

对于普通的交通出行，出行者 $i$ 的出行任务属性为{1：时间；2：金钱；3：舒适度；4：安全性；5：误点风险}，属性 1、2、5 指标越小越好；属性 3、4 越高越好；

假设：出行者 $i$ 从 C 到 D 地，地铁的旅行时间为 70 分钟，公交车的旅行时间为[40,50]，小汽车的旅行时间为[20,40]分钟；

实际结果：在该案例中，外出旅游对于准点的要求稍低，对舒适性和安全性的要求较高，出行者最终选择了使用小汽车的出行方案。

（1）应用 LT 理论

出行者 $i$ 的决策任务 2（自助外出旅游）属性集 $\mathbf{A}(D_i^2)$ 为：

$$\mathbf{A}(D_i^2) = \{1,2,3,4,5\}$$

出行者 $i$ 的决策任务 2（自助外出旅游）基本属性集 $\mathbf{A}_\alpha(D_i^2)$ 为：

$$\mathbf{A}_\alpha(D_i^2) = \{1,2,3,4,5\}$$

出行者 $i$ 的决策任务 2（自助外出旅游）占优属性集 $\mathbf{A}_\beta(D_i^2)$ 为：

$$\mathbf{A}_\beta(D_i^2) = \{1,2,3,4,5\}$$

出行者 $i$ 的决策任务 2（自助外出旅游）基本属性价值向量为：

$$\mathbf{x}(\mathbf{A}_\alpha(D_i^2)) = (\leq 60, \leq 70, \geq \text{"moderate"-3}, \geq \text{"high"-4}, \leq \text{"low"-2})$$；

出行者 $i$ 的决策任务 2（自助外出旅游）的占优属性分割为：

$$\mathbf{B}_1(D_i^2) \bigcup \mathbf{B}_2(D_i^2) = \{1,2,5\} \bigcup \{3,4\}$$

出行者 $i$ 的决策任务 2 的所有可能决策方案集合：

$$\mathbf{S}(D_i^2) = \{s_{i,1}^2 = m1, s_{i,2}^2 = m2, s_{i,3}^2 = m3\}$$

依据 2.1 小节的符号和定义，通过决策行为 $D_i^2$ 上的方案—属性映射 $f_i^2$ 得到出行者 $i$ 上决策任务 2 的基本属性价值矩阵

$$\mathbf{X}_\alpha(\mathbf{S}(D_i^2)) = (\mathbf{x}(s_{i,1}^2), \mathbf{x}(s_{i,2}^2), \mathbf{x}(s_{i,3}^2))^T = \begin{pmatrix} 70 & 50 & 3 & 5 & 1 \\ 40 \sim 50 & 50 & 3 & 4 & 2 \\ 20 \sim 40 & 60 & 4 & 4 & 2 \end{pmatrix}$$

出行者 $i$ 的决策任务 2 的占优属性价值矩阵 $\mathbf{X}_\beta(\mathbf{S}(D_i^2))$ 与基本属性价值矩阵 $\mathbf{X}_\alpha(\mathbf{S}(D_i^2))$ 完全相同；

运用 2.2 小节的 PSP 过程，比较 $\mathbf{X}_\alpha(\mathbf{S}(D_i^1))$ 的每一行和基本属性价值向量 $\mathbf{x}(\mathbf{A}_\alpha(D_i^1))$ 中每一个元素的相对优劣，由于出行方案 $s_{i,1}^2$ 的第 1 个属性的价值为 70，劣于基本属性价值向量 $\mathbf{x}(\mathbf{A}_\alpha(D_i^1))$ 第 1 个属性的基本值 60，因此出行方案 $s_{i,1}^2$ 被剔除，易验证出行方案 2 和方案 3 被保留，则出行者 $i$ 的决策任务 1 的可行决策方案集合：$\tilde{\mathbf{S}}(D_i^1) = \{s_{i,2}^2 = m2, s_{i,3}^2 = m3\}$；

再应用 2.2 小节的 LSP 过程，占优属性分割：$\mathbf{A}_\beta(D_i^2) = \mathbf{B}_1(D_i^2) \bigcup \mathbf{B}_2(D_i^2) = \{1,2,5\} \bigcup \{3,4\}$，决策行为 $D_i^2$ 的占优属性等级数 $L(D_i^m) = 2$，在最高等级 $r = 2$ 上的占优方案集合 $\hat{\mathbf{S}}(\mathbf{B}_2(D_i^2) | \tilde{\mathbf{S}}(D_i^2)) = s_{i,3}^2$（方案 $s_{i,2}^2$ 和方案 $s_{i,3}^2$ 在第 3 个属性和第 4 个属性上的价值分别为{3,4}和{4,4}，明显有$\{4,4\} \succ \{3,4\}$）。故在案例 1 中，出行者的最终出行方案为 $s_{i,3}^2$，即选择小汽车出行的方案 m3，与实际情况一致。

（2）应用 PT 理论

根据前景理论的基本假设，决策者面对获得的时候是风险厌恶的，决策者面对损失的时候是风险偏好的，由于出行行为可以看作是一种获得，因此，根据 PT 理论，决策者在选择出行方式的时候，将最小化自己的行为风险，因此将选择风险最小的出行方式，即乘地铁出行的方案 m1，与实际情况相矛盾；

（3）应用 IT 理论

由于 IT 理论的收益性测试仅仅从量的最优方面考虑行为决策，因此当方案 2 和方案 3 通过 IT 理论的兼容性测试后，无法进一步从方案 2 和方案 3 中进行选择，因此，对于本问题，IT 理论将失效。

3.3（案例 3）核电站选址决策（本案例能够用 IT、LT 理论解释，不能够运用 PT 理论解释）。

美国的核电站选址决策，实际的决策是从若干个方案中选择能够符合美国法律的地址得到候选地址，然后从几个候选地址中选择一个经济成本（价格）最低的作为最终的选址。

（1）应用 LT 理论

核电站选址决策 $D_i^3$ 的考虑的属性为：1：法律限制；2：价格。

核电站选址决策 $D_i^3$ 的属性集合：$\mathbf{A}(D_i^3) = \{1, 2\}$；

核电站选址决策 $D_i^3$ 的基本属性集合 $\mathbf{A}_\alpha(D_i^3) = \{1\}$；

核电站选址决策 $D_i^3$ 的占优属性集合 $\mathbf{A}_\beta(D_i^3) = \{2\}$

核电站选址决策 $D_i^3$ 的基本属性价值向量 $\mathbf{x}(\mathbf{A}_\alpha(D_i^3)) = (a)$；这里符号 a 表示符合美国有关的法律规定；

决策任务的占优属性分割：$\mathbf{B}_1(D_i^3) = \{2\}$（决策者在阶梯搜索阶段唯一考虑的因素为价格因素）。

依据 2.2 小节的 PSP 过程，得到可行的决策方案集合 $\tilde{\mathbf{B}}(D_i^3)$，即所有备选方案中符合美国法律的地址，然后依据 LSP 过程，决策者选择唯一占优等级（$r=1$）上的占优方案集合 $\hat{\mathbf{S}}(\mathbf{B}_1(D_i^3)|\tilde{\mathbf{S}}(D_i^3))$，即从价格角度选择价格最便宜的方案，得到最终的选址方案。

（2）应用 IT 理论

IT 理论将决策过程分为方案收集的兼容性筛选和方案选择的收益性测试两个阶段，第一阶段根据是否符合当地法律规定选择出候选地址，第二阶段对候选地址依据建设成本进行排序，选择出成本最小的地址，IT 理论对该案例的决策行为能够做出较好的解释。

（3）应用 PT 理论

PT 理论无法描述在该案例中，决策者如何收集出选址方案，又如何选择出最终的选址方案的完整过程。本案例中，对于从若干个可行的选址方案中最终确定地址,实际上不需要运用到PT理论,只需要简单地选择成本最小的方案即可，PT 理论主要关注的是决策者在决策中对于得失风险态度的刻画，对于这类不涉及到风险偏好的多属性决策问题，其分析能力受到较大局限。

3.4（案例 4）个人购物行为（本案例不能够使用 PT 理论解释，可以运用 IT、LT 理论解释）。

决策行为情景描述：某消费者在商店购买衣服，考虑的决策因素如下：1、质量；2、款式；3、颜色；4、价格；5、衣服使用时间；

实际结果：消费者对于两件看中的衣服（实际上两件衣服的款式、质量、价格完全一样，区别在于颜色和使用时间，两件衣服的颜色分别为蓝色和白色，使用时间分别为 3 年以上和 2 年以上），最终选择了可以使用 3 年以上的白色衣服。

（1）应用 LT 理论

消费者 $i$ 的决策任务 4（购物）的属性集合：$\mathbf{A}(D_i^4) = \{1, 2, 3, 4, 5\}$；

消费者 $i$ 的决策任务 4（购物）的基本属性集合：$\mathbf{A}_\alpha(D_i^4) = \{1, 2, 3, 4\}$；

消费者 $i$ 的决策任务 4（购物）的占优属性集合：$\mathbf{A}_\beta(D_i^4) = \{5\}$

消费者 $i$ 的决策任务 4（购物）的基本属性价值向量：

$\mathbf{x}(\mathbf{A}_\alpha(D_i^4)) = (\geq a, \geq b, \{red, bule, white\}, \leq 2000)$，这里，符号 a,b 分别代表质量临界值和款式临界值；

假设消费者所有可能的决策方案为衣服 $w1$ 和 $w2$，即 $\mathbf{S}(D_i^4) = \{s_{i,1}^4 = w1, s_{i,2}^4 = w2\}$，

依据 2.1 小节的符号和定义，通过决策行为 $D_i^4$ 上的方案—属性映射 $f_i^4$ 得到消费者 $i$ 的决策任务 4 的属性价值矩阵：

$$\mathbf{X}(\mathbf{S}(D_i^4)) = (\mathbf{x}(s_{i,1}^4), \mathbf{x}(s_{i,2}^4))^T = \begin{pmatrix} a & b & white & 1000 & \geq 3 \\ a & b & blue & 1000 & \geq 2 \end{pmatrix}$$

消费者 $i$ 的决策任务 4 的基本属性价值矩阵：

$$\mathbf{X}_\alpha(\mathbf{S}(D_i^4)) = (\mathbf{x}_\alpha(s_{i,1}^4), \mathbf{x}_\alpha(s_{i,2}^4))^T = \begin{pmatrix} a & b & white & 1000 \\ a & b & blue & 1000 \end{pmatrix}$$

消费者 $i$ 的决策任务 4 的占优属性价值矩阵：

$$\mathbf{X}_\beta(\mathbf{S}(D_i^4)) = (\mathbf{x}_\beta(s_{i,1}^4), \mathbf{x}_\beta(s_{i,2}^4))^T = \begin{pmatrix} \geq 3 \\ \geq 2 \end{pmatrix}$$

通过 2.2 小节的 PSP 过程，比较 $\mathbf{X}_\alpha(\mathbf{S}(D_i^4))$ 的每一行和基本属性价值向量 $\mathbf{x}(\mathbf{A}_\alpha(D_i^4))$ 中每一个元素的相对优劣，易验证两个方案都被保留，因此，消费者 $i$ 的决策任务 4 的可行决策方案集合 $\tilde{\mathbf{S}}(D_i^4) = \{s_{i,1}^4 = w1, s_{i,2}^4 = w2\}$

再应用 2.2 小节的 LSP 过程，决策行为 $D_i^4$ 的占优属性分割：$\mathbf{A}_\beta(D_i^4) = \mathbf{B}_1(D_i^4) = \{5\}$，占优属性等级数 $L(D_i^4) = 1$，在最高等级 $r = L(D_i^4) = 1$ 上的占优方案集合 $\hat{S}(\mathbf{B}_1(D_i^4)|\tilde{\mathbf{S}}(D_i^4)) = \{s_{i,1}^4\}$。故在案例 4 中，消费者最终选择方案 1，即购买白色衣服 w1，与实际情况一致。

（2）应用 PT 理论

由于本案例中没有涉及到个体的风险态度，PT 理论无法对个体如何从两件价格完全相同的衣服中最终依据穿着的时间因素做出选择给出合理的解释。

（3）应用 IT 理论

由于选择最终是根据消费者预期的衣服穿着时间给出的，衣服穿着时间可以看作一个能够量化的指标，IT 理论的收益性测试机制正好可以恰当地处理单一的定量指标，因此，本案例应用 IT 理论能够进行正确的解释。

4. 讨论

在案例 1 和案例 2 中，决策任务的情景发生了变化，一个是出差赶车，另一

个是外出旅游，虽然都涉及到决策者的风险因素，但是在第一个决策情景下，出行者对出行误点的风险更加关注，在第二个决策情景下，出行者对于出行误点的风险的重视程度下降了，取而代之的是更加关注出行中的舒适性和安全性，因此，在第1个案例中，最终的方案是选择准时到达率高的地铁方式，第2个案例中，最终选择的方案是选择更加方便舒适，同时也比较安全的小汽车出行，但在PT理论中，却无法反映出这种由于出行情景发生变化之后，出行者对于出行决策任务属性重视程度发生变化而导致最终决策行为的改变，实质原因由于PT理论是一种静态理论，较难反映决策行为的完整过程，因此，在应用该理论预测第2个案例时出现了偏差，而LT理论能够结合决策任务特征，从整个决策制定的过程进行全面考虑，分别通过PSP和LSP两个机制来刻画候选方案集的形成和最终方案的确定，因此，能够对第2个案例中出行者的最终出行方案做出正确的预测。

通过四个具体的案例对比分析，可以看到LT理论对四种常见类型的决策任务都能够给出较好的解释，而PT理论和IT理论均不能全部做出有效解释。为了更好地比较PT、IT和LT三种理论的异同，我们从对于决策过程、风险态度和多属性的处理能力三个角度对这几种理论进行了对比。

首先是决策过程，PT理论本质是属于一种静态决策理论，它主要描述决策者面临不同收益情形下对于决策风险态度如何影响到最终的决策结果，IT理论和LT理论都能够考察到完整的决策过程。其次，对于风险型决策问题的处理，PT理论和本文的LT理论都能够做出很好的解释，IT理论则没有明确地处理风险型决策问题。第三，对于决策任务的多属性特征，PT理论可以通过多参考点进行描述，但对于一般的多属性决策问题，如果决策者需要同时考虑风险型和非风险型因素，PT理论的解释力还是稍显不足，LT理论中明确地考虑到了决策任务的多属性特征，IT理论在决策的收益性测试阶段，关注的主要是单独的能够量化的属性，对于不能够量化的属性较难处理。第四，多属性感知。当面临需要考虑多个非风险型方案属性的决策情形，PT和IT理论均不能有效地描述和处理，LT理论可以很好地考虑多属性感知的问题（三、四两点，可参考案例2）。

下面的表1显示了PT、IT和LT理论之间的特点对比。其中，符号"√"表示能够胜任，符号"X"表示不能够胜任，符号"o"表示部分胜任。

表1 三种决策理论的对比结果

|    | 决策过程 | 风险态度 | 多属性描述 | 多属性感知 |
|----|------|------|-------|-------|
| PT | X    | √    | o     | X     |
| IT | √    | o    | X     | X     |
| LT | √    | √    | √     | √     |

值得注意的是，对于单方案决策，也完全可以应用本文建立的LT理论进行分析，比如，决策者面临某单个的决策方案，当该方案在最高等级上的属性子集上不满足决策者的预期标准，则决策者往往会选择放弃该方案的执行。

5. 小结

我们比较了两种典型的行为决策理论：PT理论和IT理论，PT理论由于始于对人们赌博行为的模拟，本质上还是对传统期望效用理论的改进，它的主要贡献在于在决策者的收益感知和风险态度之间建立了联系，更加符合实际地考虑了决策行为中人们在得失情形下不同的风险态度，但它最大的缺点在于没有充分考虑个体决策行为的发生过程，本质上还是通过比较个体对于风险态度差异前提下的"期望效用"进行选择，因而是一个静态的理论，对于决策行为从方案收集到最

终确定的形成机制缺乏足够的关注和有效的描述；其次，PT 理论对于多属性类型的决策问题，考虑得也不够充分；而 IT 理论的长处在于较为符合实际地描述出了个体如何进行决策的全过程，其弱点在于获得最终决策方案的收益性测试中对于多属性的特征考虑不足，其次，对于人们决策中风险态度的关注也稍为欠缺。

我们针对上述不足，提出了 LT 理论，LT 理论的特点是将决策者对多个决策属性重要程度心理感知的变化与决策方案制定过程有机地结合起来，提出了决策方案收集的初步筛选机制和最终方案确定的阶梯搜索机制，通过案例分析对比，可以发现 LT 理论能够对现实个体决策行为作出更好的解释和预测，因此是一种更普适的行为决策理论。下一步可以考虑的研究方向一是通过更多的实际决策案例或者行为实验来验证 LT 理论的合理性和有效性，另一个是以 LT 理论为基础，将其推广至多人群体决策和组织决策问题的情景中，使之发挥更大的价值。

## 参考文献

# The Ladder Theory of Behavioral Decision Making

CHEN Xing-guang

(School of Business, Jianghan University, Wuhan 430056, China)

**Abstract:** We study individual decision-making behavioral on generic view. Using a formal mathematical model, we investigate the action mechanism of decision behavioral under subjective perception changing of task attributes. Our model is built on work in two kinds classical behavioral decision making theory: "prospect theory (PT)" and "image theory (IT)". We consider subjective attributes preference of decision maker under the whole decision process. Strategies collection and selection mechanism are induced according the description of multi-attributes decision making. A novel behavioral decision-making framework named "ladder theory (LT)" is proposed. By real four cases comparing, the results shows that the LT have better explanation and prediction ability then PT and IT under some decision situations. Furthermore, we use our model to shed light on that the LT theory can cover PT and IT ideally. It is the enrichment and development for classical behavioral decision theory and, it has positive theoretical value and instructive significance for explaining plenty of real decision-making phenomena. It may facilitate our understanding of how individual decision-making performed actually.